\documentclass[conference]{IEEEtran}
\usepackage{fancyhdr}
\usepackage{graphicx,booktabs,url,amsmath}
\usepackage[caption=false]{subfig}
\usepackage{listings}

\usepackage{makecell}
\usepackage{multirow}

\begin{document}

\title{ArchAgent: Agentic AI for Heterogeneous System Design Across Chip, Scale-Up, and Scale-Out Domains}

\author{
\IEEEauthorblockN{Murali Emani}
\IEEEauthorblockA{Argonne National Laboratory, IL, USA \\ memani@anl.gov}
\and
\IEEEauthorblockN{Kashyap Manjusha}
\IEEEauthorblockA{University of Illinois Urbana-Champaign, IL, USA \\ kr58@ieee.org}

}

\maketitle
\thispagestyle{fancy}
\lhead{}
\rhead{}
\chead{}
\lfoot{\footnotesize{
SC26 Workshops, November 15-20, 2026, Chicago, Illinois, USA
\newline 979-8-3195-1221-5/26/\$31.00 \copyright 2026 IEEE}}
\rfoot{}
\cfoot{}
\renewcommand{\headrulewidth}{0pt}
\renewcommand{\footrulewidth}{0pt}

\begin{abstract}

The design space of modern AI/HPC systems has expanded rapidly across compute architectures, including GPUs, dataflow processors, and RISC-V accelerators; memory technologies, including HBM, GDDR, and SRAM; and interconnects such as NVLink, UALink, Infinity Fabric, InfiniBand, Slingshot, and Ultra Ethernet. No single architecture is optimal across LLM training, LLM serving, and AI-coupled HPC workloads, while the resulting combinatorial design space is too large for manual exploration. We present \textbf{ArchAgent}, a hierarchical multi-agent framework in which an LLM-based planner decomposes system design into chip, scale-up, and scale-out subproblems, each handled by specialist agents. A deterministic analytical evaluator closes the loop, ensuring that all performance estimates are derived from analytical models rather than generated by the LLM. On a mixed AI+HPC workload under a 1~MW power budget, ArchAgent discovers heterogeneous designs that achieve a 1.93$\times$ weighted geometric mean speedup over a homogeneous GPU baseline, while requiring only 18 evaluations compared with 100 evaluations for hill climbing, which achieves 1.65$\times$ speedup.

\end{abstract}


\section{Introduction}
The hardware substrate for large-scale AI and HPC has fragmented along three axes.
\emph{Compute} spans SIMT GPUs (NVIDIA H100/H200 and Blackwell B200/GB300; AMD Instinct MI300X/MI325X/MI355X), matrix-engine ASICs with integrated networking (AWS Trainium2~\cite{trainium}), systolic-array TPUs (Google TPU v5p, Trillium, and Ironwood~\cite{tpuv4}), SRAM-resident dataflow processors (Groq's LPU~\cite{groq}, Cerebras' wafer-scale WSE-3~\cite{cerebras}), reconfigurable dataflow units (SambaNova SN40L~\cite{sn40l}), near-memory compute architectures (d-Matrix Corsair) and RISC-V native accelerators such as Tenstorrent's Blackhole~\cite{tenstorrent}.
\emph{Memory} spans stacked HBM (HBM2e through HBM3e, with HBM4 emerging), GDDR6/GDDR7, LPDDR5X coupled to host CPUs (e.g., Grace), on-chip SRAM ranging from hundreds of megabytes (LPU) to tens of gigabytes (WSE-3), tiered SRAM--HBM--DDR hierarchies (SN40L), and CXL-attached pooled memory~\cite{cxl}; these tiers differ by orders of magnitude in capacity, bandwidth, and cost per byte.
\emph{Interconnects} have fractured into proprietary and emerging scale-up domains (NVLink, AMD Infinity Fabric, Google's ICI with optical circuit switching~\cite{tpuv4}, AWS NeuronLink, and the open UALink~\cite{ualink} standard), Ethernet-native designs that blur the scale-up/scale-out boundary (Tenstorrent's Ethernet mesh), 
and competing scale-out families (InfiniBand NDR/XDR, HPE Slingshot-11~\cite{slingshot}, RoCEv2-based fabrics such as NVIDIA Spectrum-X, and Ultra Ethernet~\cite{ultraeth}).
 Together, these choices create a heterogeneous substrate in which compute architecture, memory hierarchy, and communication fabric are increasingly co-designed rather than independently selected.


Workloads have diverged just as sharply. LLM training is dominated by GEMM and all-reduce operations, prefill is primarily compute-bound, decode is constrained by memory bandwidth, and AI-coupled HPC workloads alternate between FP64 solver phases and dense AI computation. As a result, no single chip--fabric combination performs well across all phases, motivating \emph{heterogeneous} system compositions. However, the resulting design space is highly combinatorial, spanning chip types, partition sizes, per-partition interconnects, and workload assignments. This complexity makes the space difficult to navigate through human intuition alone, while conventional black-box auto-tuners often struggle because they lack the architectural priors needed to reason about these system-level trade-offs.


LLM-based agents are well suited to this problem because they can leverage architectural priors when navigating the design space. For example, an agent can recognize that decode is typically bandwidth-bound, identify SRAM-resident dataflow architectures as a natural fit, and independently propose prefill/decode disaggregation. The risk of hallucinated performance claims can be addressed structurally: all performance estimates are produced by a deterministic analytical evaluator invoked as a tool, rather than generated by the model itself. In this design, the LLM is responsible for reasoning over architectural choices, while the evaluator provides the quantitative performance evidence used to assess them.

\begin{figure*}[t]
  \centering
  \includegraphics[width=\linewidth]{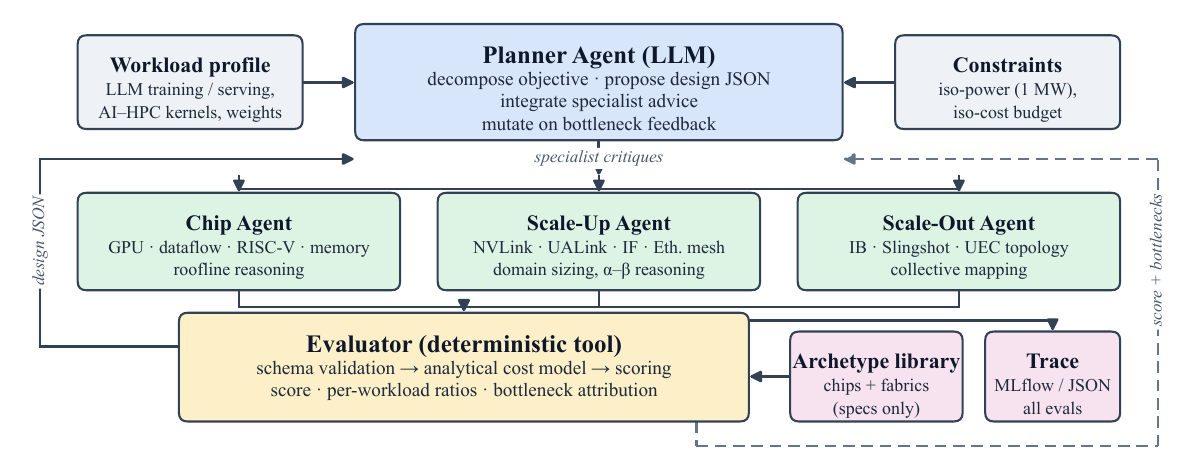}
  \caption{ArchAgent overview. The planner proposes designs as JSON; chip,
  scale-up, and scale-out sub-agents critique their subsystems; a deterministic
  evaluator validates and scores every candidate, returning bottleneck
  attribution.} 
  \label{fig:overview}
\end{figure*}


In this work, our primary contributions are: (1)~\emph{ArchAgent}, a hierarchical multi-agent framework in which a planner decomposes system design into chip, scale-up, and scale-out subproblems, which are then handled by specialist sub-agents in a closed loop with a validating evaluator. (2)~An open archetype library covering GPU, dataflow, and RISC-V compute architectures, together with ten fabric models and a calibrated analytical cost model based on a roofline chip model and $\alpha$--$\beta$ collective communication model, spanning LLM training, disaggregated serving, and four AI+HPC kernels. (3)~A mixed-workload evaluation methodology under both iso-power and iso-cost constraints, including a same-generation control that isolates the benefits of heterogeneity from those of process-generation improvements. (4)~A systematic evaluation showing that agent-designed heterogeneous systems can outperform homogeneous procurement, complemented by search-efficiency and ablation studies that isolate the contributions of the agent hierarchy and evaluator feedback.

\section{Related Work}


\paragraph{Architecture DSE}
Prior work has extensively explored automated accelerator design-space exploration. ArchGym~\cite{archgym} uses RL, GAMMA~\cite{gamma} employs genetic search, and AutoDSE~\cite{autodse} applies bottleneck-guided search; related systems such as Timeloop/Accelergy~\cite{timeloop} and MAESTRO~\cite{maestro} explore accelerator mappings and dataflows. These approaches generally treat the evaluator as a black box and optimize within a predefined parameter space, without explicitly reasoning about architectural relationships that transfer across workloads. \emph{ArchAgent}, in contrast, uses an LLM to exploit architectural priors while coordinating system-level search.

\paragraph{LLMs for hardware}
Recent work applies LLMs to hardware development, including ChipNeMo~\cite{chipnemo} for chip-design workflows and SpecLLM~\cite{specllm} and RTL-oriented systems~\cite{rtllm,rtlcoder} for specification and RTL generation. These efforts primarily target individual chips or design artifacts; we instead consider full-system composition, jointly selecting compute architectures, scale-up fabrics, and scale-out networks.

\paragraph{Performance modeling}
Our evaluator follows the analytical modeling tradition of Calculon~\cite{calculon} and LLMCompass~\cite{llmcompass} for transformer performance and ASTRA-sim~\cite{astrasim2} for hierarchical collective communication. We abstract intra-chip mapping, addressed by Timeloop/Accelergy~\cite{timeloop,accelergy}, through effective utilization factors and focus on system-level performance across heterogeneous compute and interconnects.


To our knowledge, \emph{ArchAgent} is the first framework to pair hierarchical LLM agents with a deterministic evaluator to jointly explore chip, scale-up, and scale-out design, auditable configurations together with the reasoning behind key design choices.

\section{The ArchAgent Framework}

\subsection{Design Representation}

\subsection{Design Representation}
We represent each candidate system as a JSON document containing a list of \emph{partitions}. Each partition specifies a chip archetype, a scale-up fabric, a scale-out fabric, a fraction of the facility power budget, and a workload \emph{role} (\texttt{general}, \texttt{prefill}, \texttt{decode}, \texttt{solver}, or \texttt{ai}). Workload roles allow the evaluator to model disaggregated execution without requiring the agent to explicitly manage placement. For example, a serving configuration can assign prefill to a GPU partition and decode to an SRAM-resident dataflow partition. Before any performance evaluation, a schema validator checks the design for basic feasibility, including fabric bandwidth that exceeds the chip's available port capacity, power allocations that exceed the facility budget, and insufficient FP64 capability for HPC workloads. Validation errors are returned to the agent as structured feedback, preventing evaluation cycles from being spent on invalid or unsupported configurations.

\subsection{Agent Hierarchy}

The \emph{planner} agent proposes and iteratively revises complete system designs. For each candidate, it must invoke the \texttt{evaluate\_design} tool, which provides the planner with the hardware catalog, design schema, and optimization objective, but not with predicted performance improvements. Three \emph{specialist sub-agents} provide additional guidance during the search: a chip agent reasons about compute archetypes, roofline constraints, and memory hierarchies; a scale-up agent applies $\alpha$--$\beta$ reasoning to intra-domain communication and domain sizing; and a scale-out agent analyzes collective communication and system topology. Their recommendations are returned as concise parameter-level guidance and attached to the corresponding evaluation result.
The \emph{evaluator} is a deterministic analytical model. Given a valid configuration, it returns the mixed-workload score, per-workload performance relative to the reference system, and an attribution of the dominant bottleneck among compute throughput, memory bandwidth, fabric latency, and bisection bandwidth. A key structural invariant is that the LLM agents never generate or assert performance estimates themselves; all performance quantities originate from the deterministic evaluator. This separation makes the search process auditable: the complete sequence of candidate designs, evaluation results, and agent recommendations can be recorded and inspected.


\subsection{Analytical Cost Model}
We model conventional accelerator performance using a roofline formulation with dtype-specific peak FLOP rates, machine-utilization factors (MFU), and memory-bandwidth efficiency factors. For dataflow architectures whose sharded weights fit within on-chip SRAM, we instead model pipelined layer sharding, accounting for per-stage computation and point-to-point activation transfers rather than tensor-parallel all-reduce. This captures the execution model of SRAM-resident architectures such as Groq and Cerebras.

Communication costs are modeled using the standard $\alpha$--$\beta$ formulation~\cite{thakur}. The model supports ring and hierarchical all-reduce, with communication first within a domain and then across domains, as well as all-to-all and point-to-point transfers. For LLM training, we combine tensor, pipeline, and data parallelism (TP/PP/DP)\cite{megatron}, including pipeline-bubble overhead and a ZeRO-style\cite{zero} memory-feasibility check. For serving, we model disaggregated prefill and decode~\cite{distserve} and assign each phase to partitions according to their workload roles.

The HPC suite consists of four workloads chosen to exercise distinct compute and communication regimes: a 27-point FP64 stencil dominated by halo exchange, an HPCG-like conjugate-gradient (CG) kernel with latency-sensitive scalar all-reduce, a 3D FFT dominated by all-to-all communication and global bisection bandwidth, and a coupled AI-surrogate--solver loop in which successive phases can be pipelined across heterogeneous partitions. Section~\ref{sec:e1} describes calibration of the analytical model against published MLPerf results~\cite{mlperf} and vendor-reported measurements.

\section{Evaluation}
\subsection{Methodology}

All systems are provisioned under an \emph{iso-power} constraint of 1,MW at the facility level, including the power consumed by fabric ports. Results under an \emph{iso-cost} constraint, based on publicly available list prices, are reported as a secondary analysis. We optimize a weighted geometric mean of per-workload speedups relative to a homogeneous H100$+$NVLink4$+$InfiniBand-NDR reference system, whose score is normalized to 1.0. The workload weights are ${0.25, 0.30, 0.10, 0.10, 0.10, 0.15}$ for training, serving, stencil, HPCG, FFT, and the coupled workload, respectively, representing a mixed AI+HPC facility profile. A configuration receives a score of zero if any workload cannot be executed.

To isolate the benefits of architectural heterogeneity from those of newer process generations, the headline comparison restricts all candidates to a common hardware generation. This constraint is necessary because allowing unrestricted generational mixing can make the optimization trivial: in preliminary searches, the optimizer selected B200 accelerators throughout, with the performance advantage attributable primarily to the newer generation rather than to heterogeneous system design. Random search and hill climbing are evaluated under the same number of evaluator calls as ArchAgent.

\subsection{Model Calibration}\label{sec:e1}

An automated verification suite comprising 32 checks validates the model at three levels: specification invariants against vendor datasheets, asymptotic identities for collective communication models, and end-to-end performance predictions against published measurements. As representative calibration points, the model predicts 910 tok/s/GPU for BF16 training of Llama-70B on a 1,024-H100 cluster, consistent with the approximately 750--1,100 tok/s/GPU reported by MLPerf-class runs~\cite{mlperf}. For FP8 decode at TP4 with batch size 32, the model predicts 109 tok/s/user, within the approximately 60--150 tok/s/user range reported for TRT-LLM. For HPCG, the model produces iteration times on the millisecond scale, consistent with the memory-bandwidth-limited behavior reported for comparable systems.

\subsection{Heterogeneous vs Homogeneous (iso-power)}
\begin{table}[t]
\caption{Best mixed-workload score by search method (1\,MW iso-power;
reference: homogeneous H100$+$NVLink4$+$IB-NDR $=1.000$).}
\label{tab:e2}
\centering\small
\begin{tabular}{llrrr}
\toprule
Method & Planner LLM & Evals & Runs & Score (mean$\pm$sd) \\
\midrule
ArchAgent & gpt-oss-120b & 100 & 1 & \textbf{1.93} \\
ArchAgent & gpt-oss-120b & 50 & 2 & 1.71$\pm$0.16 \\
ArchAgent & gemma-4-31B-it & 50 & 3 & 0.87$\pm$0.15 \\
Hill climb & --- & 100 & 1 & 1.65 \\
Hill climb & --- & 50 & 1 & 1.61 \\
Random & --- & 100 & 1 & 1.51 \\
\bottomrule
\end{tabular}
\end{table}

\begin{figure}[t]
  \centering
    \includegraphics[width=0.92\columnwidth]{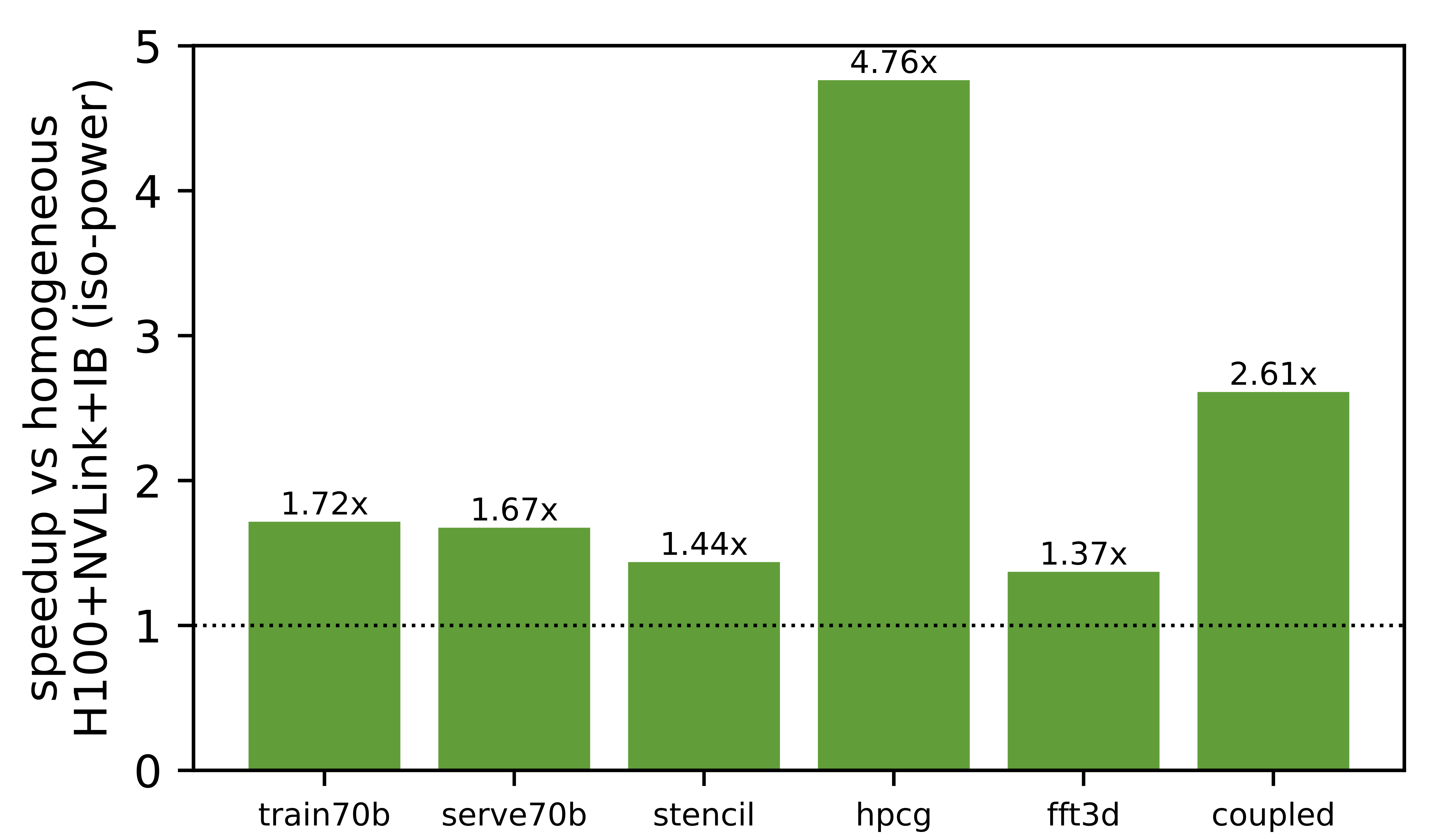}
  \caption{Per-workload speedup of the best agent-discovered design
  (score 1.93) over the homogeneous reference at iso-power.}
  \label{fig:workloads}
\end{figure}


Table~\ref{tab:e2} summarizes design quality across search methods and planner models. The best design identified by ArchAgent (Fig.\ref{fig:workloads}) 
emerged from the planner's response to evaluator-reported bottlenecks rather than from random exploration. The resulting configuration improves performance across all workloads, achieving 1.72$\times$ for training, 1.67$\times$ for serving, 4.76$\times$ for HPCG, and 2.61$\times$ for the coupled AI$+$HPC workload; FFT, at 1.37$\times$, is the binding constraint.

Planner capability also affects search outcomes. Under identical evaluation budgets and guardrails, gemma-4-31B-it converges to a score of 0.87, below the homogeneous reference, and produces 7 schema-invalid and 4 otherwise infeasible designs in 50 evaluations. In comparison, gpt-oss-120b produces 5 schema-invalid and no infeasible designs under the same conditions. These results suggest that the framework can also serve as a controlled probe of how model capability affects architectural search, although the observed differences reflect the complete agent-and-model system rather than model capability alone. The hardware used to host the planner 
affects only search wall-clock time, as expected given the deterministic evaluator.

Finally, the reported designs benefit in part from the generational advantage of B200 over the H100 reference. The same-generation control experiment, reported in the artifact, isolates the contribution of architectural heterogeneity from this generational effect.

\subsection{Search Efficiency}
\begin{figure}[t]
  \centering
  \includegraphics[width=0.92\columnwidth]{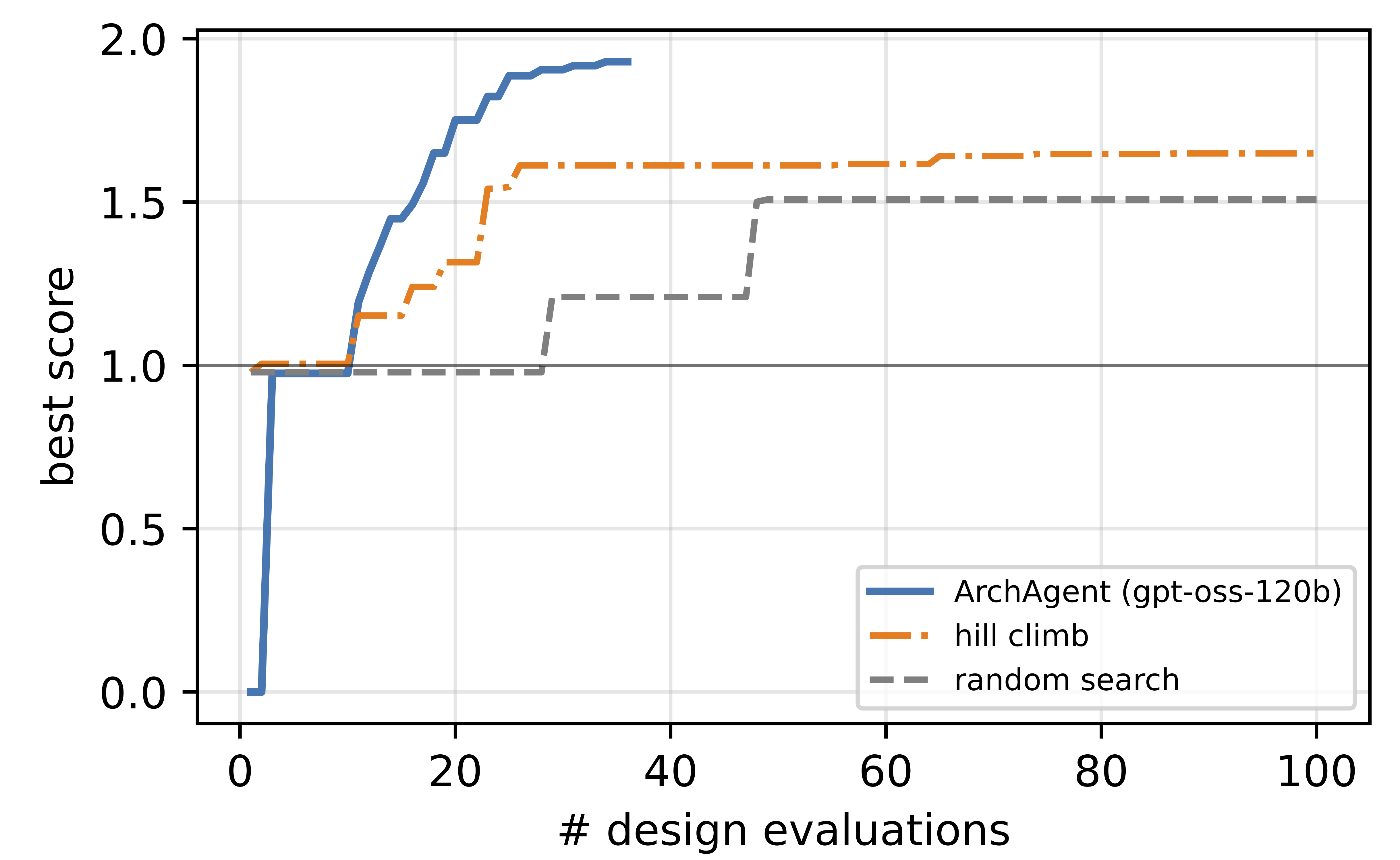}
  \caption{Best score vs.\ evaluations (100-eval budget): ArchAgent
  (gpt-oss-120b) vs.\ hill climb and random search.}
  \label{fig:e3}
\end{figure}


Fig.~\ref{fig:e3} shows convergence under a matched 100-evaluation budget. ArchAgent surpasses hill climbing's final score of 1.65 by evaluation~18 and reaches its best score of 1.93 by approximately evaluation~36, after which no further improvement is observed. Random search requires 48 evaluations to exceed a score of 1.5. The faster convergence is consistent with the architectural structure encoded in the agent hierarchy: the planner's early proposals already incorporate multi-partition configurations and workload-role assignments, whereas hill climbing must discover these structural choices through local mutations.

\subsection{Implications for Future Systems}

The remaining performance gaps in the winning configurations consistently concentrate along two axes: all-to-all communication and FP64 memory bandwidth. In particular, FFT performance remains constrained by global bisection bandwidth, while the stencil workload is limited by FP64 memory throughput. These results suggest three directions for future mixed AI+HPC systems. (i)~Scale-out fabrics would benefit from higher injection bandwidth, with $\geq$200GB/s per endpoint as a representative target, together with hardware support for in-network collective and all-to-all communication. (ii)~Memory systems with HBM4-class bandwidth ($\geq$10TB/s per socket), or accelerators that combine high memory bandwidth with substantial FP64 capability through dataflow or near-memory compute, could reduce the performance gap on bandwidth-bound HPC workloads. (iii)~System procurement could be organized around workload roles rather than a single system. 
Across our search results, the optimizer frequently allocates 2--10\% of the power budget to specialized decode or solver partitions, suggesting that a relatively small heterogeneous component can complement a predominantly AI-oriented system and account for a substantial portion of the improvement over homogeneous configurations.

\section{Discussion and Limitations}

Our conclusions are bounded by the fidelity of the analytical evaluator. The current model does not capture switch-level congestion, uses fixed factors for compute/communication overlap, and does not yet account for KV-cache transfer in disaggregated serving. These simplifications can affect the relative evaluation of communication-intensive configurations. The wafer-scale archetype is approximated as a set of composable slices, which may underestimate intra-wafer bandwidth and overestimate inter-slice communication performance. Similarly, the iso-cost analysis inherits uncertainty from publicly reported list prices.

These limitations concern the evaluator rather than the framework: the agent loop is evaluator-agnostic, allowing the analytical model to be replaced or augmented by higher-fidelity models. As a next step, we plan to validate selected top-ranked configurations with simulator-in-the-loop evaluation, using ASTRA-sim~\cite{astrasim2} for collective communication and Timeloop~\cite{timeloop}-class tools for chip-level mapping, while keeping such expensive evaluations outside the primary search loop.
On the agent side, LLM stochasticity motivates reporting results across multiple random seeds. Search outcomes also depend on planner capability. We view this dependence as an experimentally useful property of the framework: because performance estimates are generated exclusively by the deterministic evaluator, differences in planning capability can be reflected directly in the quality of the resulting configurations under a fixed evaluation environment. This makes ArchAgent a potential benchmark for studying open-weight models on hardware-design search tasks. Ablations isolating the contributions of the specialist-agent hierarchy and bottleneck-attribution feedback remain future work. 

\section{Conclusion}



In this paper, we present ArchAgent, a framework that combines hierarchical LLM agents with a deterministic evaluator to explore heterogeneous system designs. Our results show that this approach can search complex architectural design spaces more efficiently than stochastic baselines while identifying higher-quality configurations. Under a mixed AI+HPC workload profile and an iso-power constraint, ArchAgent achieves a 1.93$\times$ improvement over the homogeneous GPU reference and reaches a design quality exceeding hill climbing's final result within 18 evaluations.

The configurations discovered by ArchAgent also exhibit a consistent architectural pattern: a large GPU-like partition complemented by smaller, role-specialized dataflow and FP64 partitions. This result suggests that future computing facilities may benefit from composing hardware around workload roles rather than relying exclusively on homogeneous accelerator fleets. At the component level, the remaining performance bottlenecks highlight two capabilities that are less prominent in current AI-oriented accelerators: global bisection bandwidth and FP64 memory bandwidth. These results point to both capabilities as promising directions for future accelerator and system design.


\section{Acknowledgement}

Argonne National Laboratory's work was supported by the U.S. Department of Energy, Advanced Scientific Computing Research (ASCR), under contract DE-AC02-06CH11357. 
Claude was used to generate Figure~1 and to assist with refining the manuscript text.

\bibliographystyle{IEEEtran}
\bibliography{refs}

\appendix
\subsection{Discovered System Configurations}\label{app:configs}
We report the best design at each level of the pipeline: the agent's design
artifact, (the exact JSON it emitted, reproducible from the released trace),
the deployment specification implied by iso-power provisioning, and the
\emph{realized workload mapping} that is, the placement and parallelism the
evaluator selects when executing each workload on the design. All numbers
below are for the best gpt-oss-120b run (100-evaluation budget, score 1.93).

\subsection{Design Artifact (agent output)}

The agent's final design artifact is shown below. It specifies the hardware
partitions, their roles, power allocations, and interconnects.

\begin{lstlisting}[basicstyle=\ttfamily\scriptsize,
columns=fullflexible]
{"name":"Iter33",
"partitions":[
{"chip":"B200","role":"general","power_frac":.98,
"scaleup":"NVLink5","scaleout":"IB-XDR"},
{"chip":"WSE3s","role":"decode","power_frac":.02,
"scaleup":"UALink","scaleout":"UEC"}]}
\end{lstlisting}

\subsection{Deployment Specification (1,MW iso-power)}

Under the 1 MW iso-power constraint, the design yields 896 B200
accelerators organized into twelve NVL72 domains and four WSE-3 slices in a
single UALink domain. The B200 and WSE-3 partitions use InfiniBand XDR and
Ultra Ethernet 800, respectively, for scale-out. Table~\ref{tab:bom}
summarizes the resulting deployment. The role labels are used as inputs to
workload mapping but do not impose hard placement constraints.

\begin{table}[h]
\caption{Design proposed} 
\label{tab:bom}
\centering\scriptsize
\begin{tabular}{@{}lll@{}}
\toprule
 & Partition 1 & Partition 2 \\
\midrule
Role (agent label) & general & decode \\
Chip & B200 $\times$ 896 & WSE-3 slice $\times$ 4 \\
Scale-up & NVLink5, 72-GPU domains & UALink, 4-slice domain \\
 & (12 NVL72 racks) & (1 node) \\
Scale-out & InfiniBand XDR & Ultra Ethernet 800 \\
 & 100\,GB/s/endpoint & 100\,GB/s/endpoint \\
\bottomrule
\end{tabular}
\end{table}

\subsection{Realized Workload Mapping}
The evaluator maps each workload to the best role-eligible partition and
searches parallelism per workload; Table~\ref{tab:mapping} shows the
resulting execution configuration i.e., what an operator would actually
deploy.

\begin{table}[h]
\caption{Execution configuration selected by the evaluator for each
workload on design \texttt{Iter33} (Llama-70B for LLM workloads).}
\label{tab:mapping}
\centering\scriptsize
\begin{tabular}{@{}llll@{}}
\toprule
Workload & Partition & Configuration & Result \\
\midrule
Training & B200 & best TP/PP/DP mapping, & 1.72$\times$ ref.\\
 & & global batch 1024 & \\
Serving: prefill & B200 & TP1, batch 8, & 12.96M tok/s \\
 & & 896 replica groups & \\
Serving: decode & B200 & TP1, batch 128, & 4.71M tok/s \\
 & & 896 replicas, FP8, & (41 tok/s/user, \\
 & & & mem-BW bound) \\
Stencil / HPCG / FFT & B200 & FP64, all 896 GPUs & 1.44 / 4.76 / 1.37$\times$ \\
Coupled AI$+$HPC: & B200 & FP64 stencil phase & \multirow{2}{*}{2.61$\times$ (pipelined)} \\
\quad solver phase & & & \\
Coupled AI$+$HPC: & WSE-3 & FNO surrogate, & \\
\quad AI phase & & SRAM-resident & \\
\bottomrule
\end{tabular}
\end{table}


\subsection{Role Labels vs.\ Realized Placement}

The role assigned by the agent does not necessarily determine the partition
used during execution. In \texttt{Iter33}, the wafer-scale partition is
labeled \texttt{decode}, but the evaluator assigns both serving stages to
the B200 partition. With only four wafer slices available, the 70B FP8
model cannot be accommodated in on-slice SRAM: sharding requires
35,GB per slice, compared with 5.5,GB of available SRAM. Consequently,
SRAM-resident decode is not feasible at this budget, and the evaluator
selects the B200 partition, which provides the required capacity through
HBM3e.

The WSE-3 partition is instead used for the \emph{AI-surrogate phase} of
the coupled AI$+$HPC workload. For the FNO proxy, its dense FP16
throughput is higher than the effective throughput of the B200 partition,
and pipelining the AI and solver phases produces a 2.61$\times$ coupled
speedup.

Thus, role labels serve as guidance to the workload mapper rather than as
hard placement constraints. The realized mapping reported in
Table~\ref{tab:mapping} reflects the actual execution configuration, while
both the agent-specified roles and the resulting placement can be
reconstructed from the audit trail. At larger power budgets
($\geq 8$ wafer slices per replica), SRAM-resident decode becomes feasible,
causing the \texttt{decode} role and the realized placement to coincide.

\end{document}